\documentstyle[12pt]{article}

\begin{document}
\pagestyle{empty}
$\ $
\vskip 1.5 truecm

\centerline{\bf
Zero Modes on Linked Strings
}

\

\centerline {Jaume Garriga}
\smallskip
\centerline
{\it
IFAE,  Edifici C, Universitat Aut\`onoma de Barcelona
}
\centerline
{\it
E-08193 Bellaterra, Spain.
}

\

\centerline
{Tanmay Vachaspati
\footnote{Address from 1 January 1995:
Department of Physics, Case Western Reserve University,
Cleveland, OH 44106, USA.}
}
\smallskip
\centerline
{\it
Isaac Newton Institute, 20 Clarkson Road,}
\centerline
{\it
University of Cambridge, Cambridge, CB3 0EH, U.K.
}

\vskip 1. truecm

\begin{abstract}

We study linked loops of string in the presence of
bosonic condensates and fermionic zero modes on the strings.
We find that the strings necessarily carry a current if the
bosons have an Aharanov-Bohm interaction with the string. The
fermionic case is analyzed in the context of the standard model
where there are lepton and quark zero modes on $Z-$strings.
Here we find that the fermionic ground state in the linked
string background is lower than the ground state when the
loops are unlinked but otherwise identical. As in the bosonic case,
the $Z-$strings carry a non-vanishing
electric current in the ground state. The baryon number
of the linked configuration is found to agree with
previous indirect results. We also evaluate the angular
momentum, electromagnetic charge and baryonic three current
on the linked $Z-$string configuration.
Finally we point out a possible gravitational analogue
of the linked string system.

\end{abstract}

\clearpage
\pagestyle{plain}

\section{Introduction}

The study of particle-soliton interactions over the last 20 years
has led to the discovery of novel physical phenomena (for a few
such examples, see \cite{rjcr,jgfw,wi85}).
Some of these phenomena have been successfully tested
in condensed matter systems where solitons abound and the
ideas have also been widely applied to hypothetical
particle physics situations.
These beautiful ideas are also relevant to the standard
model of
particle physics even though the model does not contain any
{\it topological}
defects. This is because the ideas  have applicability
even when the solitons are not truly topological and there are
a number of such ``non-topological'' solitons present in the standard
model. In this paper we shall apply some of these ideas to
the electroweak $Z-$string - a flux tube containing magnetic
flux of the $Z$ boson \cite{yn,nm,tv1}.

Several facets of the $Z-$string have been studied over the past
few years and recently the attention has turned to their interactions
with the particles in the standard model. Earnshaw and Perkins\cite{mewp}
have found the quark and lepton zero modes
on a straight $Z-$string. This means that $Z-$strings are superconducting
and can carry of order $10^8$ Amperes of electric current. In
another paper it was shown that linked loops of $Z-$string carry
baryon number\cite{tvgf}. This result was derived by
integrating out the anomaly equation for the baryonic current
and did not explicitly consider the fermions interacting with the string.
The moment seems ripe for a more direct
computation of the baryon number and other charges on linked loops
of $Z-$string. This is the underlying motivation for the present
work.

The technical analysis given in this paper is inspired by
earlier work in which the spectral flow in non-trivial
backgrounds was studied. In particular,
the paper by Manton on Schwinger electrodynamics in 1+1
dimensions\cite{manton}
was immensely helpful as the situation there closely resembles
the problem at hand.

The plan of the paper is as follows. In section 2, we set up our
conventions, describe electroweak $Z-$strings and
recapitulate known results on fermionic zero modes.
In section 3 we describe two cases in which the Dirac equation
for the fermionic zero modes can be solved completely in terms
of the string profile functions. In both cases we need to
assume the Bogomolnyi limit\cite{bogo} in which the masses of the
scalar and vector fields making up the string are equal.
In the first of the two cases,
the fermion is massless (such as the neutrino)
while in the second case we have a ``super-Bogomolnyi'' limit in
which the fermion mass is equal to the scalar and vector masses
and the $Z-$charge on the left-handed fermion vanishes.
Then, in section 4, we
describe what happens to bosonic zero modes on linked strings as
this is the simplest situation. The main results of this paper
are found in section
5, where we discuss lepton and quark zero modes on linked loops
of electroweak $Z-$string. Here we evaluate the energy, angular
momentum and the electromagnetic and baryonic currents on the
strings. One outcome of this evaluation is that the
ground state energy of fermions on linked strings is lower than
that on unlinked strings. We also confirm previous findings
that linked loops of $Z-$string carry baryon number
$2N_F cos(2\theta_W)$ where $N_F$ is the number of fermion families
and $\theta_W$ is the Weinberg angle. In addition, we
show that the linked strings necessarily carry
electromagnetic currents in their ground state. We extend our
analysis to loops with linkage greater than one and
plot the energy as a function of linkage. We conclude in
section 6 and outline a gravitational situation that resembles
the system of linked strings and which may have physics similar
to that discussed in this paper.

\

\

\

\section{Review}

The bosonic sector of the standard model of the electroweak interactions
describes an $SU(2)\times U(1)$ invariant theory with a scalar field
$\Phi$ in the fundamental representation of $SU(2)$. It
is described by the Lagrangian:
\begin{equation}
L_b = L_W + L_Y + L_{\Phi} - V(\Phi )
\label{2.1}
\end{equation}B
where,
\begin{equation}
L_W = - {1 \over 4} W_{\mu \nu a} W^{\mu \nu a}
\label{2.2}
\end{equation}
\begin{equation}
L_Y = - {1 \over 4} Y_{ \mu \nu} Y^{\mu \nu}
\label{2.3}
\end{equation}
where $W_{\mu \nu}^a$ and $Y_{\mu \nu}$ are the field
strengths for the $SU(2)$ and $U(1)$ gauge fields $W_\mu ^a$ and $Y_\mu$
respectively. Also,
\begin{equation}
L_\Phi = |D_\lambda \Phi |^2 \equiv
   \biggl |\biggl (\partial _\lambda -
             {{ig} \over 2} \tau ^a W_\lambda ^a -
                  {{ig'} \over 2} Y_\lambda \biggr) \Phi \biggr| ^2
\label{2.4}
\end{equation}
\begin{equation}
V(\Phi ) = \lambda (\Phi ^{\dag} \Phi - \eta ^2 /2 )^2 \ ,
\label{2.5}
\end{equation}
where,
\begin{equation}
\Phi = \pmatrix{\phi^+\cr \phi\cr}
\end{equation}
is a complex doublet.

In addition, the fermionic sector Lagrangian is:
\begin{equation}
L_f = L_l + L_q
\label{2.6}
\end{equation}
where, the lepton and quark sector Lagrangians for a single family are:
\begin{equation}
L_l = - i {\bar \Psi} \gamma^\mu D_\mu \Psi
      - i {\bar e}_R \gamma^\mu D_\mu e_R
      + h({\bar e}_R \Phi^{\dag} \Psi + {\bar \Psi} \Phi e_R )
\label{2.7}
\end{equation}
\begin{eqnarray}
L_q =& -i ({\bar u} , {\bar d})_L \gamma^\mu D_\mu
         \pmatrix{u\cr d\cr}_L
      -i {\bar u}_R \gamma^\mu D_\mu u_R
      -i {\bar d}_R \gamma^\mu D_\mu d_R
\\
&
-G_d \biggl [ ({\bar u}, {\bar d})_L \pmatrix{\phi^+\cr \phi\cr} d_R
+{\bar d}_R (\phi^{-} , {\phi^*}) \pmatrix{u\cr d\cr}_L \biggr ]
\\
&
 - G_u \biggl [ ({\bar u}, {\bar d})_L \pmatrix{-{\phi^*}\cr
 \phi^{-}\cr} u_R + {\bar u}_R (-\phi, \phi^+) \pmatrix{u\cr d\cr}_L
       \biggr ]
\\
\label{2.8}
\end{eqnarray}
with $\phi^- = ( \phi^+ )^*$.
The indices $L$ and $R$ refer to left- and right-handed components.

In our analysis we will only be dealing with one fermion family at
a time and hence we shall not be considering the effects of family
mixing such as occurs due to the KM matrix. For our purpose, this
simplification is justified because the string configuration has
vanishing charged currents. It is known that in the neutral current
sector the gauge and mass matrices can be diagonalized
simultaneously\cite{chengli} and hence there are no complications
involving family mixing\footnote{The {\it decay} of $Z-$strings
necessarily involves the charged current sector and we expect family
mixing effects (such as CP violation) to be present in such processes.}.

The covariant derivatives occuring in the electroweak Lagrangian
are:
\begin{equation}
D_\mu \Psi = D_\mu \pmatrix{\nu \cr e\cr}_L =
\biggl ( \partial_\mu - {{ig} \over 2} \tau^a W_\mu ^a +
{{ig'} \over 2}  Y_\mu \biggr ) \pmatrix{\nu \cr e\cr}_L
\label{DPsi}
\end{equation}
\begin{equation}
D_\mu e_R = ( \partial_\mu + ig' Y_\mu ) e_R
\label{Der}
\end{equation}
\begin{equation}
D_\mu \pmatrix{u\cr d\cr}_L =
\biggl ( \partial_\mu - {{ig} \over 2} \tau^a W_\mu ^a -
{{ig'} \over 6}  Y_\mu \biggr ) \pmatrix{u\cr d\cr}_L
\label{Dul}
\end{equation}
\begin{equation}
D_\mu u_R = ( \partial_\mu - {{i2g'} \over 3} Y_\mu ) u_R
\label{Dur}
\end{equation}
\begin{equation}
D_\mu d_R = ( \partial_\mu + {{ig'} \over 3} Y_\mu ) d_R
\label{Ddr}
\end{equation}
The conventions used here are those in Ref. \cite{chengli}.

We also need to define
\begin{equation}
Z_\mu \equiv cos\theta_W n^a {W_\mu} ^a - sin\theta_W Y_\mu \ ,
\ \ \ \
A_\mu \equiv sin\theta_W n^a {W_\mu} ^a + cos\theta_W Y_\mu \ ,
\label{2.9}
\end{equation}
where,
$n^a$ is the unit vector
\begin{equation}
n^a = - {{ \Phi^{\dag} \tau^a \Phi} \over {\Phi^{\dag} \Phi}}
\label{2.10}
\end{equation}
and,
\begin{equation}
tan\theta_W \equiv {{g'} \over g} \ , \ \ \ \
\alpha \equiv (g^2 + {g'} ^2 )^{1/2} \ .
\label{2.11}
\end{equation}

For our purposes, only the $Z$ gauge field is non-zero and all
the other gauge fields can be set to zero. This effectively
reduces the bosonic part of the Lagrangian to an Abelian-Higgs
model.

There are two distinct string solutions to the bosonic sector
of the model - the $W$ and $Z$ strings\cite{tvmb,mbtvmb}.
Here we shall only consider
the unit winding $Z$ string for which the solution is:
\begin{equation}
\Phi = {\eta \over {\sqrt{2}}} f(r) e^{i\theta}
\pmatrix{0\cr 1\cr} \ ,
\ \ \
Z_\theta = - {2 \over \alpha} {{v(r)} \over {r}}
\label{2.12}
\end{equation}
with all other fields set to zero. The solution in eqn. (\ref{2.12})
is given
in cylindrical coordinates $(r, \theta , z)$ and the profile
functions $f$ and $v$ satisfy:
\begin{equation}
f'' + {{f'} \over r} - {f \over {r^2}} (1-v)^2 +
\lambda \eta^2 (1-f^2) f = 0
\label{2.13}
\end{equation}
\begin{equation}
v'' - {{v'} \over r} + {{\alpha^2 \eta^2} \over  4} f^2 (1-v) = 0
\label{2.14}
\end{equation}
\begin{equation}
f(0)=v(0)=0 \ , \ \ \ f(\infty )=v(\infty ) =1
\label{2.15}
\end{equation}
Note that in writing eqn. (\ref{2.13}) we are
only considering the bosonic sector and have set the fermionic
fields to zero.

The total $Z$ magnetic flux in a string can
be evaluated directly from eqns. (\ref{2.12}) and (\ref{2.15}):
\begin{equation}
F_Z = {{4\pi} \over \alpha} \ .
\label{2.16}
\end{equation}
This means that any particle whose $Z-$charge is not an integer
multiple of $\alpha /2$ will have an Aharanov-Bohm interaction
with the $Z-$string.

In Ref. \cite{tvgf}, the baryon number on linked strings (Fig. 1) was
calculated by integrating out the anomalous baryonic current conservation
equation:
\begin{equation}
\partial_\mu j^\mu _{B} = {{N_F} \over {32\pi^2}}
 [ -g^2 W^a _{\mu \nu} {\tilde W}^{a \mu \nu} +
                         {g'}{}^2 Y_{\mu \nu} {\tilde Y}^{\mu \nu} ].
\label{2.17}
\end{equation}
where, $N_F$ is the number of families and ${\tilde W}^{a \mu \nu}
\equiv (1/2) \epsilon^{\mu \nu \lambda \sigma} W^a _{\lambda \sigma}$.
The result is\footnote{We have included the factor of 2 error in
Ref. \cite{tvgf} that is reported in the errata.}:
\begin{equation}
Q_B = 2 N_F cos(2\theta_w ) \
\label{2.18}
\end{equation}
up to the addition of an integer.
This calculation is indirect because it makes no explicit reference to
the fermions. In Sec. 5 we shall recover (\ref{2.18})
directly by studying the fermionic zero modes on the string.

The Dirac equations for a single family of leptons and quarks
have been solved in the background of a straight
$Z$ string by Earnshaw and Perkins\cite{mewp}. The representation
for the $\gamma$ matrices they use is:
\begin{equation}
\gamma^r = \pmatrix{0&e^{-i\theta}&0&0\cr
                    -e^{i\theta}&0&0&0\cr
                    0&0&0& -e^{-i\theta}\cr
                    0&0&e^{i\theta}&0\cr}\ , \ \ \
\gamma^\theta = \pmatrix{0&-ie^{-i\theta}&0&0\cr
                         -ie^{i\theta}&0&0&0\cr
                         0&0&0&ie^{-i\theta}\cr
                         0&0&ie^{i\theta}&0\cr}\ ,
\label{gamma1}
\end{equation}
\begin{equation}
\gamma^0 = \pmatrix{\tau^3&0\cr 0&-\tau^3\cr}\ , \ \ \
\gamma^z = \pmatrix{0&{\bf 1}\cr -{\bf 1}&0\cr}\ , \ \ \
\gamma^5 = \pmatrix{0&{\bf 1}\cr {\bf 1}&0\cr}\ .
\label{gamma2}
\end{equation}
For the electron they find
the zero mode solution\footnote{We have corrected an
error of a minus sign in eqn. (19) of Ref. \cite{mewp}.}:
\begin{equation}
e_L = \pmatrix{1\cr 0\cr -1\cr 0\cr} \psi_1 (r) \ , \ \ \
e_R = \pmatrix{0\cr 1\cr 0\cr 1\cr} i \psi_4(r)
\label{2.19}
\end{equation}
where,
\begin{equation}
\psi_1 ' + {{qv} \over r} \psi_1 = -h{\eta \over \sqrt{2}}
f \psi_4
\label{2.20a}
\end{equation}
\begin{equation}
\psi_4 ' - {{(q-1)v} \over r} \psi_4 = -h {\eta \over \sqrt{2}}
 f \psi_1 \ .
\label{2.20b}
\end{equation}
$q$ denotes the $Z$ charge of the various left-handed
fermions and for the electron we have
$q = cos(2\theta_W)$. The boundary conditions are
that $\psi_1$ and $\psi_4$ should vanish asymptotically.
This means that there is only one arbitrary constant of
integration in the solution to eqns. (\ref{2.20a}) and
(\ref{2.20b}). This may be taken
to be a normalization of $\psi_1$ and $\psi_4$.

Earnshaw and Perkins have also solved the Dirac equations for the
quark zero modes. For the $d$ quark, the solution is the same
as in eqns. (\ref{2.19}), (\ref{2.20a}) and (\ref{2.20b}) except that
$q=1 -  (2/3)sin^2\theta_W$. For the $u$ quark
the solution is:
\begin{equation}
u_L = \pmatrix{0\cr 1\cr 0\cr -1\cr} \psi_2 (r) \ , \ \ \
u_R = \pmatrix{1\cr 0\cr 1\cr 0\cr} i \psi_3(r)
\label{2.21}
\end{equation}
where,
\begin{equation}
\psi_2 ' - {{qv} \over r} \psi_2 = -G_u {\eta \over \sqrt{2}}
f \psi_3
\label{2.22a}
\end{equation}
\begin{equation}
\psi_3 ' + {{(q+1)v} \over r} \psi_3 = -G_u {\eta \over \sqrt{2}}
f \psi_2
\label{2.22b}
\end{equation}
where, $q = -1 + (4/3) sin^2\theta_W$. Note that
(\ref{2.20a}), (\ref{2.20b}) are related to (\ref{2.22a}),
(\ref{2.22b}) by $q \rightarrow -q$.

The neutrino zero modes can be found explicitly in terms of
the string profile functions since the neutrino is massless and
the right-hand sides of the neutrino Dirac equations (corresponding
to eqns. (\ref{2.22a}) and (\ref{2.22b})) vanish.
This is done in the next section.

We are mainly interested in the situation where two
loops of string are linked (Fig. 1). To analyse this
situation, we consider two circular loops of string
of winding number $n$ that are linked but the strings
themselves are far from each other.  The loops
are taken to be very much larger than the thickness of
the strings and we shall always be working to lowest order
in the ratio $w/a$ where $w$ is the string thickness and
$a$ is the radius of the circular loops. In this approximation,
we can ignore the curvature of the loop and consider the
strings as being straight on scales much larger than
$w$ but much smaller than $a$.
Then the effect of one loop on the
other is to produce a winding Higgs field and a winding gauge field
with zero field strength at the location of the second loop.
We can write down the field configuration of the linked
loops in terms of the Higgs and gauge fields for two large
loops in this approximation. If $(\phi^{(1)}, Z_\mu^{(1)})$
and $(\phi^{(2)}, Z_\mu^{(2)})$ are the fields for
two isolated loops then the fields for the full configuration are
given by the product ansatz:
\begin{equation}
\phi = {{\sqrt{2}} \over \eta} \phi^{(1)} \phi^{(2)} \ , \ \ \
Z_\mu = Z_\mu^{(1)} + Z_\mu^{(2)} \ .
\label{linkstring1}
\end{equation}
If we set up a local coordinate frame on loop 1, with the
$z$ axis along the string, the fields in the
vicinity of the string can be found by using (\ref{2.12}):
\begin{equation}
\phi =  \phi^{(1)} e^{i n z/a} \ , \ \ \
Z_\mu = Z_\mu^{(1)} - {{2n} \over {\alpha a}} {\hat e}_z \ .
\label{linkstring2}
\end{equation}

\

\

\

\section{Fermionic Zero Mode Solutions}

In this section we give explicit solutions to the
Dirac equations in (\ref{2.20a}) and (\ref{2.20b})
in two special cases. The first
case is when the fermion is massless and the second
case is when the fermion mass is equal to the scalar
mass which is also equal to the vector mass. The first
of these is relevant to neutrino zero modes. The second
is not relevant to the electroweak case but may be of
interest in other situations.

We start out by writing eqns. (\ref{2.20a}) and
(\ref{2.20b}) in more compact form by defining:
\begin{equation}
F_1 = {\rm exp}
\biggl [+q \int dr {v \over r} \biggr ] \psi_1 \ ,
\ \ \
F_4 = {\rm exp}
\biggl [-(q-1) \int dr {v \over r} \biggr ] \psi_4 \ .
\label{3.1}
\end{equation}
Then,
\begin{equation}
F_1 ' = -{{h\eta f} \over \sqrt{2}} {\rm exp}
\biggl [+(2q-1) \int dr {v \over r} \biggr ] F_4  \ ,
\ \ \
F_4 ' = -{{h\eta f}\over \sqrt{2}} {\rm exp}
\biggl [-(2q-1) \int dr {v \over r} \biggr ] F_1  \ .
\label{3.2}
\end{equation}

For arbitrary parameters, the integral occuring in (\ref{3.2})
is not known. However, in the Bogomolnyi limit - when the
Higgs mass ($m_H = \sqrt{2\lambda }\eta$) and the $Z$ boson
mass ($m_Z = \alpha \eta /2$) are equal -
we can perform the integration. This is because, in this case,
the equations for the string are equivalent to the following
first order differential equations\cite{bogo}:
\begin{equation}
f' = {f \over r} (1-v)
\label{3.4a}
\end{equation}
\begin{equation}
v' =  {{m_Z^2} \over 2} r (1-f^2 ) \ .
\label{3.4b}
\end{equation}
Now eqn. (\ref{3.4a}) yields:
\begin{equation}
\int dr {v \over r} = {\rm ln} \biggl ( {m_Z r \over f} \biggr )
\label{vr}
\end{equation}
where we have included a factor of $m_Z$
to make the argument of the logarithm dimensionless.

Therefore, when the fermion is massless ($h=0$), the solutions
are simply
\begin{equation}
\psi _1 = c_1 m_Z^{3/2} \biggl ( {{m_Z r} \over f} \biggr )^{-q}
\  ,  \ \ \
\psi _4 = c_4 m_Z^{3/2} \biggl ( {{m_Z r} \over f} \biggr )^{q-1}
\label{3.3}
\end{equation}
where, $c_1$ and $c_4$ are independent constants that can be
chosen to normalize the left- and right-handed fermion states
and the spinors are given in (\ref{2.19}).
The boundary condition that the left-handed
fermion wavefunction should vanish at infinity is only
satisfied if $q > 0$. Hence (\ref{3.3}) can only give a valid
solution for $q > 0$ for the left-handed fermion. If we also
require normalizability, we
need $q > 1$. (Note that there is no singularity at
$r=0$ because $f \propto r$ when $r \sim 0$.) If we have
a left-handed fermion with
$q \le -1$, the correct equations to use are the equations
corresponding to the up quark equations given in
(\ref{2.22a}) and (\ref{2.22b}) and these are solved
by letting $q \rightarrow -q$ in (\ref{3.3}). In this
case, the spinors are given in (\ref{2.21}).

For the neutrino, the right-handed component is absent
and $q = -1$. This means that the neutrino has the same
spinor structure as the left-handed up quark and the solution
is that in (\ref{3.3}) with $q$ replaced by $+1$.
Therefore the wave function falls off as $1/r$
and the state is strictly not normalizable - the normalization
integral diverges logarithmically. However, depending on the
physical situation, one could be justified in imposing
a cut-off. We shall be considering closed loops of string
and the cutoff in our case is given by the radius of the
loop.

Next, we consider the ``super-Bogomolnyi'' limit when the mass
of the fermion ($m_f = h\eta /\sqrt{2}$)
is also equal to the Higgs and $Z$ boson masses.
In this case, when the charge
on the left-handed fermion vanishes ($q=0$),
by staring at the eqns. (\ref{3.2}) with (\ref{vr}),
(\ref{3.4a}) and (\ref{3.4b}) we guessed the solution to be:
\begin{equation}
\psi _1 = N m_Z ^{3/2} (1-f^2)
\label{3.5a}
\end{equation}
\begin{equation}
\psi _4 =
2 N m_Z^{1/2}  {f \over r} (1-v)
\label{3.5b}
\end{equation}
where $N$ is a dimensionless normalization factor.

The solution for the up quark equations can be written down
by using the transformation $q \rightarrow -q$ in the above
solutions.

The solution in (\ref{3.5a}) and (\ref{3.5b}) can probably also
be derived using supersymmetry arguments such as described by
di Vecchia and Ferrara \cite{pvsf} but we are unaware of such
a derivation in the literature.

The left-handed fermion wave-functions found above can be multiplied
by a phase factor ${\rm exp}[i(E_p t - pz)$ and the resulting
wave-function will still solve the Dirac equations provided
\begin{equation}
E_p = \epsilon_i p
\label{ep}
\end{equation}
where, $i$ labels the fermions, and,
\begin{equation}
\epsilon_\nu = +1 = \epsilon_u \ , \ \ \
\epsilon_e   = -1 = \epsilon_d \ .
\label{epsi1}
\end{equation}
In other words,
$\nu_L$ and $u$ travel parallel to the string flux while
$e$ and $d$ travel anti-parallel to the string flux.

\

\

\

\section{Bosonic Condensates on Linked Strings}

Before proceeding to the fermionic case, let us consider the simpler
case of bosonic superconductivity.  If A and B are two loops of string
linked to each other, then we shall see that
there is a current induced in loop A
provided that the bosonic condensate within its core has fractional charge
with respect to the ``magnetic" fl\-ux  trapped in B.

The model we consider is the original
Witten model \cite{wi85} modified in order to allow for the
Aharonov-Bohm-type  interaction. We have two complex scalar fields
$\Sigma$ and $\Phi$ and two U(1) gauge fields $A_{\mu}$ and $B_{\mu}$. The
Lagrangian is
\begin{equation}
{\cal L}=|D_{\mu}\Phi|^2+|D_{\mu}\Sigma|^2-
{1\over 4}A_{\mu\nu}A^{\mu\nu}-{1\over 4}B_{\mu\nu}B^{\mu\nu}
-{\cal V}(|\Phi|,|\Sigma|).
\label{bosonlagrangian}
\end{equation}
Here, $D_{\mu}\Phi=(\partial_{\mu}+i(q/2)B_{\mu})\Phi$,
$D_{\mu}\Sigma=(\partial_{\mu}+i(e/2)A_{\mu}+i(q'/2)B_{\mu})\Sigma$,
$A_{\mu\nu}=\partial_{\mu}A_{\nu}-\partial_{\nu}A_{\mu}$ and
$B_{\mu\nu}=\partial_{\mu}B_{\nu}-\partial_{\nu}B_{\mu}$.
This Lagrangian is invariant under $U(1)\times U(1)$
gauge transformations:
\begin{equation}
\Sigma\to \Sigma e^{-iq'\Lambda},\quad \Phi\to \Phi e^{-iq\Lambda},
\quad B_{\mu}\to B_{\mu}+2\Lambda,_{\mu},
\label{gaugetransformations}
\end{equation}
and:
\begin{equation}
\Sigma\to \Sigma e^{-ie\tilde\Lambda},
\quad A_{\mu}\to A_{\mu}+2\tilde\Lambda,_{\mu},\label{em}
\end{equation}
where $\Lambda$ and $\tilde\Lambda$ are arbitary functions.
The potential
\begin{equation}
{\cal V}=
\lambda_{\phi}(|\Phi|^2-\eta^2)^2+f(|\Phi|^2-\eta^2)|\Sigma|^2+
\lambda_{\sigma}|\Sigma|^4+m^2|\Sigma|^2,
\end{equation}
is chosen so that the field $\Phi$ acquires a vacuum expectation
value. Since the vacuum manifold is non-trivial, $\Phi$ will admit
string-like solutions.
The field $A_{\mu}$ is then identified with electromagnetism, which is
unbroken outside the string, and $\Sigma$ plays the role of a charge carrier.

Straight string solutions can be studied in cylindrical coordinates using the
ansatz
\begin{equation}
\Phi=\phi(r) e^{in\theta},\quad B=B_{\theta}(r) d\theta,\quad
\Sigma=\sigma(r), \quad A=A_{\theta}(r) d\theta,
\label{ansatz1}
\end{equation}
where $n$ is an integer (the winding number). The field equations then take the
form
\begin{equation}
\phi''+{\phi'\over r}-{1\over r^2} (n+{q\over 2}
B_{\theta})^2\phi={1\over 2}{\cal
V},_{\phi},\label{phi}
\end{equation}
\begin{equation}
\sigma''+{\sigma'\over r}-{1\over r^2}
({e\over 2}A_{\theta}+{q'\over 2}
B_{\theta})^2\sigma={1\over 2}{\cal V},_{\sigma},
\end{equation}
\begin{equation}
B''_{\theta}-{B'_{\theta}\over r}=q \phi^2(n+{q\over 2}B_{\theta})+
q'\sigma^2({e\over 2}A_{\theta}+{q'\over 2}B_{\theta}),\label{btheta}
\end{equation}
\begin{equation}
A''_{\theta}-{A'_{\theta}\over r}=
e\sigma^2({e\over 2}A_{\theta}+{q'\over 2}B_{\theta}).\label{solenoid}
\end{equation}
With $\sigma=0$, these equations admit the usual Nielsen-Olesen
vortex solutions \cite{hnpo}.
The function $\phi(r)$ vanishes for $r\to 0$ and it
approaches its vacuum expectation value $\eta$ exponentially fast for
$r>\delta$, where $\delta\sim\lambda_{\phi}^{-1/2}\eta^{-1}$ is the thickness
of the string core. The gauge field outside the core is given by
\begin{equation}B_{\theta}={-2n/ q},\label{asy}\end{equation}
so that $D_{\mu}\Phi=0$ for $r>>\delta$.  Therefore the
fl\-ux of the $B_{\mu}$ field is quantized,
\begin{equation}
F_B=4\pi n/q, \label{bflux}
\end{equation}
and this will be important later on.

Although $\sigma=0$ is always a solution, Witten \cite{wi85} showed
that there is a range of parameters for which this solution
is unstable.  For
$f\eta^2>m^2>0$ the effective mass squared for $\sigma$ is negative in the
region where $\phi=0$, so that a condensate $\sigma\neq 0$ typically forms in
the core of the  string.
To introduce a current along the string
we need to consider ``low energy'' excitations on top of the condensate, of the
form
\begin{equation}
\Sigma=\sigma(r)e^{i\psi(z,t)}.\label{phase}
\end{equation}
In the Lorentz
gauge, the equation of motion for $\Sigma$ splits
into a real and an imaginary part
$$
\Delta \sigma-V_{\mu}V^{\mu}\sigma={1\over 2}{\cal V},_{\sigma}
$$
\begin{equation}
\psi,_{tt}-\psi,_{zz}=0,\label{boxpsi}
\end{equation}
where $V_{\mu}\equiv \psi,_{\mu}+{e\over 2} A_{\mu}+{q'\over 2}
B_{\mu}$ and $\Delta$ is the
transverse Laplacian. The electromagnetic current on the string
is given by
\begin{equation}
J_{\mu}
={\partial {\cal L}\over \partial A^{\mu}}=
e\sigma^2(\psi,_{\mu}+{e\over 2}A_{\mu}+{q'\over 2}B_{\mu}).
\label{current}
\end{equation}
Note that the phase $\psi$ obeys a massless wave
equation in the longitudinal space, whose general solution is
$\psi=f(t+z)+g(t-z)$, with $f$ and $g$  arbitrary functions. However, if we
are interested in exact solutions within a simple ansatz (see below),  we
have to restrict attention to one of the following special cases: $\psi\propto
t$,
corresponding to a constant charge per unit length, $\psi\propto
z$, corresponding to a constant space-like current, or one of the forms
$\psi= f(t+z)$, $\psi= g(t-z)$, corresponding to light-like currents of
arbitrary profile propagating along the string \cite{gape94}.
More general forms for $\psi$
will radiate and settle down to
one of these special cases. For our problem, the
relevant configuration will be the one with a
space-like current, $\psi\propto z$.

The reason why linked loops must carry non-vanishing currents
can be seen from Eq.(\ref{current}). Neglecting, for the time being,
the contribution of $A_{\mu}$,
the circulation of $J_{\mu}$ along one of the loops is
proportional to
$$
\oint \psi,_{\mu}dx^{\mu}+{q'\over 2}\oint B_{\mu}dx^{\mu}=\Delta \psi+
{2\pi q'\over q},
$$
where we have used that the circulation of $B_{\mu}$ along one
loop is equal to the flux (\ref{bflux}) trapped in the core of the other loop.
Since $\Sigma$ has to be single valued, $\Delta \psi$
must be a multiple of $2\pi$ . Therefore
$J_{\mu}$ cannot vanish unless $q'/q$ is an integer.

Let us now  make a more quantitative estimate of the actual current induced
on the loops. As mentioned in section 2, to lowest order in $w/a$,
we can mimick the linkage by taking just one straight string along the z-axis
with appropriate periodic boundary conditions.
To this end, we generalize the ansatz
(\ref{ansatz1}), allowing for a twist in the $\Phi$ field (see
(\ref{linkstring2})):
\begin{equation}
\Phi=\phi(r) e^{in\theta}e^{inz/a}.
\end{equation}
Identifying $z=-\pi a$ with $z=\pi a$, this will represent a
loop of radius  $a$ and winding number $n$ which is being threaded by another
string of winding number $n$, as in Fig. 1.

In the presence of a $z$ dependent
phase in  (\ref{phase}), it is  necessary to introduce longitudinal
components $A_{z}(r)$ and $B_z(r)$ for the gauge fields. Then, in
addition to eqns. (\ref{btheta}), (\ref{solenoid}) and
(\ref{boxpsi}) above, we have
$$
B''_z+{1\over r}B'_z=q\phi^2({n\over a}+{q\over 2} B_z)+
{q'\over 2}\sigma^2 V_z,
$$
\begin{equation}
A''_z+{1\over r}A'_z= e \sigma^2 V_z
\label{az}
\end{equation}
$$
\phi''+{\phi'\over r}-\left[({n\over a}+{q\over 2}B_z)^2+{1\over
r^2}(n+{q\over 2}B_{\theta})^2\right]\phi={1\over 2}{\cal V},_{\phi}.
$$
As with the $\theta$ component, the condition that $D_z\Phi$ vanishes outside
the core of the string yields
the asymptotic value $B_z=-2n/(a q)$. This represents the pure
gauge winding around the `other' string.

The current flowing along the string will be given, from (\ref{current}), by
\begin{equation}
J= Ke(\psi,_z+ {e\over 2} \bar A_z+ {q'\over 2} \bar B_z).\label{j1}
\end{equation}
Here $K\equiv \int d^2x \sigma^2$, where the integral is over the transverse
section of the string, $\psi=lz/a$, where $l$ is an integer, and $\bar A_{z},
\bar B_{z}$ are the gauge fields averaged over the cross section of the core
with weight given by $\sigma^2$. To lowest order in the gauge couplings, $\bar
B_{z}$ is given by its asymptotic value $-2n/(aq)$, as indicated above.
Following Witten \cite{wi85}, in order to estimate $\bar A_z$ we consider
the formal solution of Eq. (\ref{az}). For an
idealized infinitely thin current we have
\begin{equation}
A_z({\bf x})=- {1\over 4\pi}
\int dz {1\over |{\bf x}-{\bf x}(z)|} J(z). \label{formal}
\end{equation}
The logarithmic divergence in the right hand side of (\ref{formal})
can be regularized by  considering a current of
finite thickness $m_{\sigma}^{-1}$. Then, along the string
$$\bar A_z \approx -{1\over 2\pi} \ln(m_{\sigma} a) J,$$
where $J$ is given by (\ref{j1}). Eliminating $\bar A_z$ from both equations we
have (to lowest order in gauge couplings)
$$
J\approx {K e\over a[1+K(e^2/4\pi) \ln(m_{\sigma}a)]}
\left(l-n{q'\over q}\right),
$$
And, neglecting unity in front of the logarithmic term,
\begin{equation}
\bar A_z\approx {4\pi\over ae}\left(l-n{q'\over q}\right).
\label{a1}
\end{equation}
Therefore, if $nq'/q$ is not an integer, we reach the conclusion that
there has to be at least a current of magnitude $J\sim (ae\ln
[m_{\sigma}a])^{-1}$ flowing along the string.

A new interesting phenomenon can be seen
if we go beyond lowest order in $e$.
Note that for $q'\neq 0$ there has to be a
magnetic field
$A_{r\theta}\neq 0$  trapped in the core of the string, since Eq.
(\ref{solenoid}) has a source term of the form $e\sigma^2q'B_{\theta}$. This
source term
corresponds to a solenoid-like current that wraps around the outer layers of
the string core. Since the thickness of the condensate is of
order $(m_{\sigma})^{-1}\equiv(\eta^2 f-m^2)^{-1/2} {\buildrel >\over \sim}
\delta$, and the asymptotic value of $B_{\theta}$ is given by (\ref{asy}),
the magnetic fl\-ux of $A_{\mu}$ trapped within the ``solenoid'' will be of
order
\begin{equation}
F_A\sim en{q'\over q}{\sigma_0^2\over m_{\sigma}^2}, \label{aflux}
\end{equation}
where $\sigma_0\sim\lambda^{-1/2}m_{\sigma}$ is the magnitude of the
condensate. Depending on the parameters, this flux can be quite large and have
interesting consequences.

In particular, when two
loops are linked it is not possible to have the circulation of $A_{\mu}$
vanish along the loops, and a current will arise. This effect is of
order $e^2/\lambda_{\sigma}$ compared to the lowest order
contribution (\ref{a1}), but can give the dominant effect if $nq'/q$ is an
integer (in which case we can set $l=nq'/q$), or if the self coupling
$\lambda_{\sigma}$ is sufficiently small.

\

\

\

\section{Fermionic Zero Modes on Linked Strings}

The Dirac equations that we need to solve for the electron
and $d$ quark are of the general form:
\begin{equation}
i\gamma^\mu D_\mu ^{(q)} \psi_L = h \phi \psi_R \ ,
\ \ \
i\gamma^\mu D_\mu ^{(q-1)} \psi_R = h \phi^* \psi_L \ .
\label{dirac}
\end{equation}
where, $D_\mu ^{(q)} = \partial_\mu + i q Z_\mu$
and the bosonic background is given in (\ref{linkstring2}).
For the $u$ quark, $\phi$ must be replaced
by $\phi^*$ and $q-1$ by $q+1$ in the second equation.
For the neutrino we should set $h=0$.

The Dirac equations (\ref{dirac}) split up into equations
in the transverse directions to the string and in the
longitudinal direction. In what follows, we will only
be considering the zero mode solutions to the transverse
part of the Dirac equation as these are the low energy
modes that live on the string and are the dominant
new effect due to the string background. The other massive
modes would also feel the string background but we expect
the effects due to these modes to be smaller.
A rigorous analysis, however,
would have to take all the transverse modes, including the
massive ones, into account.

To solve eqn. (\ref{dirac}), we consider the ansatz
\begin{equation}
\psi_L = e^{- i(E_p t - p z)} \psi_L^{(0)}(r) \ , \ \ \
\psi_R = e^{- i(E_p t - (p-n/a) z)} \psi_R^{(0)}(r) \
\label{ansatz}
\end{equation}
where, the superscript $(0)$ denotes that the function
is the one found on the straight string as described in
sections 2 and 3 and $n$ denotes the winding number of the
two strings (assumed equal). The coordinates
$(t,r,\theta ,z)$ are
cylindrical coordinates with the string locally along the
$z$ axis and we are confining our attention to the vicinity
of a point on one of the strings. The functions
$\psi_L^{(0)}$ and $\psi_R^{(0)}$ should be normalized
so that
\begin{equation}
\int d^3 x \psi^{\dag} \psi =
\int d^3 x (|\psi _L^{(0)}|^2+|\psi _R^{(0)}|^2 )
= 1
\label{normalization}
\end{equation}
This condition completely fixes the fermion wavefunction
since there is only one arbitrary normalization factor in
the solution. (For example, see eqns. (\ref{3.5a}) and
(\ref{3.5b}).) But without an explicit solution, we
cannot say what the normalization of the
left-handed component is relative to the right-handed component.

To lowest order in $w/a$,
eqn. (\ref{ansatz}) is a solution provided $E$ and $p$
satisfy the dispersion  relation:
\begin{equation}
E_p = \epsilon_i ( p + q Z_z ) \
\label{dispersion}
\end{equation}
where, the symbol $\epsilon_i$ is defined below eqn. (\ref{ep}),
$Z_z$ is given by (\ref{linkstring2}) and $ap$ has to be
an integer for single-valuedness of the wavefunctions.
So, in terms of $\omega \equiv aE$ and
$k \equiv a p \in \cal{Z}$, we have,
\begin{equation}
\omega_k = \epsilon_i  ( k - qZ  ) \ .
\label{omega}
\end{equation}
where,
\begin{equation}
Z \equiv {{2 n} \over \alpha}
\label{zna}
\end{equation}
The crucial property of this equation is that, if there
is an Aharanov-Bohm interaction between the $Z-$string
and the fermion, $\omega_k$ cannot be zero for any value
of $k$ since $k$ is an integer.

The energy of the fermions is found by summing over the
negative frequencies  - that is, the Dirac sea - and so
the energy $E$ is:
\begin{equation}
E = {1 \over a} \sum_{\omega_k < 0} \omega_k
  = \epsilon_i {1 \over a} \sum_{\omega_k < 0}  ( k - q Z )
  = \epsilon_i {1 \over a} \sum_{k=k_F}^{-\epsilon_i  \infty}
                       ( k - q Z )
\label{energy}
\end{equation}
where, $k_F$ denotes the Fermi level - the value of $k$ for
the highest filled state. Therefore we need to sum a series
of the type:
\begin{equation}
S = \sum_{k=k_F}^{\infty} ( k - q Z )
  = \sum_{k=0}^{\infty} ( k+k_F - q Z ) \ .
\label{series}
\end{equation}
The sum is found using the zeta function
regularization\cite{grry} :
\begin{equation}
S = \zeta (-1, k_F -q Z) =
 - {1 \over {12}} - {1 \over 2} (k_F-qZ)(k_F-q Z-1)
\label{sum}
\end{equation}
With this result, the energy contribution from the fermions
takes the form:
\begin{equation}
E_i = -{1 \over {24a}} +
{1 \over {2a}} \biggl [
      k_F^{(i)} - q_i Z + {\epsilon_i \over 2}
               \biggr ] ^2
\equiv -{1 \over {24a}} + {1 \over {2a}} K_i ^2
\label{ei}
\end{equation}
where, the index $i$ is a label for the particular fermion
in question. As the non-trivial
term\footnote{The contribution $1/12a$ is the Casimir
energy since the strings are closed and is present whether
the loops are linked or unlinked.} in $E_i$ comes
in the form of a square, the ground state can be found by
minimizing $E_i$ with respect to $k_F^{(i)}$ for each $i$.
The result of this exercise is summarized in Table 1
where we give the charges, the Fermi levels
$k_F^{(i)}$ (for $\gamma = sin^2 \theta_W$ close to
$1/4$) and the value of the minimum energy contribution
for each fermion $(\nu_L, e,d,u)$ from one family.
(The other families contribute identically.)
If we now also
include the fact that the quarks come in 3 colours
and that we have 2 loops that are linked, the total energy is:
\begin{equation}
E = 2N_F ( E_\nu + E_e + 3E_d + 3 E_u ) =
    {{4N_F} \over {3a}} [ 8 \gamma^2 - 6 \gamma + 1 ] .
\label{totale}
\end{equation}

Note that $E_i$ attains its maximum value of $1/12a$ if
$q_i Z$ is integral
and its minimum value of $-1/24a$ if $q_i Z$ is half integral.
When the loops are unlinked, we have $q_i Z = 0$ and so
the ground state energy of unlinked loops has to be larger
than or equal to the ground state energy of linked loops.
An explicit calculation along the lines used for linked loops
(eqn. (\ref{totale})) for two unlinked loops yields:
\begin{equation}
E ({\rm unlinked }) = {{4N_F} \over {3a}}
\label{eunlinked}
\end{equation}
and so the difference of the unlinked and linked ground state
energies is:
\begin{equation}
\Delta E =  {{8N_F} \over {3a}} \gamma [4\gamma -3] \ .
\label{de}
\end{equation}
Inserting physical values $\gamma = 0.23$ and $N_F = 3$,
we get,
\begin{equation}
\Delta E = -3.8/a \ .
\label{diffe}
\end{equation}

Note that in calculating the energy (\ref{totale}),
(\ref{eunlinked}) we have included the contribution from the
neutrinos by using the formula (\ref{ei}). As the neutrino
is massless and the zero mode state is only normalizable on
imposing a cut-off, it is an open question if it is legitimate
to ignore the contribution of the other neutrino modes besides
the zero mode. This is a subtle issue which has been discussed
in the context of 2+1 dimensions in the existing literature
\cite{hdv}. In the present context, we are considering closed
string loops in 3+1 dimensions and the issue deserves further
investigation. Here we will only remark that the question
applies regardless of whether the loops are linked or not and
so the issue has no bearing on the difference in energies
between linked and unlinked loops given in (\ref{diffe}).

The energy momentum for fermions can be found by varying the
action with respect to the metric and is:
\begin{equation}
T_{\mu \nu} = {i \over 2}  [
{\bar \psi} \gamma_{(\mu} D_{\nu )} \psi -
(D_{(\mu } \psi )^{\dag} \gamma^0 \gamma_{\nu )} \psi ] \ .
\label{tmn}
\end{equation}
This leads to the result that the magnitude of the fermion
momentum along the string is equal to the energy of the fermion
- a result which could also be deduced by noting that the
fermions are massless on the string. Now to find the
total momentum, we should recall that the electron
and $d$ quark travel in one direction while the neutrino and
$u$ quark travel in the opposite direction. Using this fact,
we find that the magnitude of the angular momentum due to
the fermionic ground state on one of the strings is:
\begin{equation}
\Omega = a N_F (E_\nu - E_e - 3 E_d + 3 E_u ) = 0
\label{om}
\end{equation}
when the linkage is one.

In non-trivial but symmetric backgrounds, it is possible to define
a generalized angular momentum operator for the fluctuations on
top of the background. If we confine our attention
to the case where we have a loop in the $xy-$plane and $n$ straight
infinite strings along the $z-$axis threading the circular loop,
this system has rotational symmetry about the $z-$axis. The operator
that annihilates the field configuration is the generalized angular
momentum operator and is \cite{mjlptv}
\begin{equation}
M_z = L_z + S_z + n I_z
\label{gam}
\end{equation}
where,
\begin{equation}
L_z = -i {\bf 1} {{\partial \ } \over {\partial \theta}} \ ,
\label{lz}
\end{equation}
$S_z$ is the spin operator, and, the isospin operator is given in
terms of the $U(1)$ (hypercharge) and $SU(2)$ charges - $q_1$ and
$q_2$ respectively - of the field in question:
\begin{equation}
I_z = {1 \over 2} \biggl [
      \biggl ( {{2q_2} \over g} \biggr ) \tau^3 -
      \biggl ( {{2q_1} \over {g'}} \biggr ) {\bf 1}
                  \biggr ] \ .
\label{iz}
\end{equation}
The isopin operator acts via a commutator bracket on the gauge fields
and by ordinary matrix multiplication on the Higgs field and fermion
doublets.

We are interested in the angular momentum of the fermions on the
circular loop which lies entirely in the $xy-$plane. The fermions in
the zero modes therefore have $S_z =0$. The action of $L_z$ is found
by acting on the fermion wave-functions such as in (\ref{ansatz})
(remembering to let $n \rightarrow -n$ for the neutrino and up quark).
The action of $I_z$ is found by using the charges of the fermions
given in (\ref{DPsi})-(\ref{Ddr}). We then find:
\begin{equation}
M_z \pmatrix{\nu_L\cr e_L\cr} =
  \pmatrix{(k^{(\nu )} +n)\nu_L \cr k^{(e)} e_L\cr} \ , \ \ \
M_z \pmatrix{u_L\cr d_L\cr} =
  \pmatrix{(k^{(u)} +{n\over 3})u_L \cr
           (k^{(d)} -{{2n}\over 3}) d_L\cr} \ ,
\label{mzpsil}
\end{equation}
\begin{equation}
M_z e_R = k^{(e)} e_R \ , \ \
M_z u_R = (k^{(u)} +{n\over 3}) u_R \ , \ \
M_z d_R = (k^{(d)} - {{2n}\over 3}) d_R \
\label{mzul}
\end{equation}
where the $k^{(i)}$ are defined above eqn. (\ref{omega}).
Now summing over states, as in the case of the energy, we find the
total generalized angular momentum of the fermions on the circular
loop:
\begin{equation}
{\cal M} =
  {1\over 2} \biggl [ k_F ^{(\nu )} +n +{1\over 2} \biggr ] ^2 -
 {1\over 2} \biggl [ k_F ^{(e)}-{1\over 2} \biggr ] -
{3\over 2} \biggl [ k_F ^{(d)}-{{2n}\over 3} -{1\over 2}\biggr ] ^2 +
{3\over 2} \biggl [ k_F ^{(u)}+{{n}\over 3} +{1\over 2}\biggr ] ^2 \ .
\label{totgam}
\end{equation}
Note that though the gauge fields do not enter explicitly in the
generalized angular momentum, they do play a role in determining the angular
momentum of the ground state through the values of the Fermi levels.
If we now consider the case $n=1$, and use the values of the Fermi
levels from Table 1, we find ${\cal M} = 0$.

We now calculate the electromagnetic and baryonic charges and
currents on the linked loops. For this we must sum over the
charges of each filled state. Since the states up to the state
$k_F^{(i)}$ are filled, we need to find
a sum of the kind:
\begin{equation}
S_q = \sum_{k=k_F}^{ \infty} 1 \ .
\label{sq1}
\end{equation}
To regularize the divergence of the series, we write it as
\begin{equation}
S_q =\lim_{\lambda\to 0} \sum_{k=k_F}^{ \infty} (k-q Z)^{\lambda} \ .
\label{sq2}
\end{equation}
Note that we have chosen to use the gauge invariant combination
$k-qZ$ rather than $k$ or some other gauge non-invariant
expression\cite{manton}.
Now we can use the zeta function regularization to get:
\begin{equation}
S_q =  \sum_{k=0}^{\infty} (k+k_F -q Z)^0  =
 \zeta(0, k_F-q Z ) =
 - \biggl [ k_F -qZ - {1 \over 2} \biggr ] \ .
\label{sqdone}
\end{equation}
With this result, we find the contribution to the charge due
to fermion $i$:
\begin{equation}
Q_i = \epsilon_i q_i
      \biggl [ k_F^{(i)} -q_i Z + {\epsilon_i \over 2} \biggr ]
     = \epsilon_i q_i K_i
\label{qi}
\end{equation}
The currents along the $z$
direction are given by ${\bar \psi} \gamma^z \psi$ where
$\gamma^z$ is given in eqn. (\ref{gamma2}). This gives
\begin{equation}
J_i = \epsilon_i Q_i \ .
\label{ji}
\end{equation}

Now we can evaluate the electric and baryon charges and currents
for each of the fermions $e$, $d$ and $u$. The results are
exhibited in Table 1. Finally we can find the total charges
and currents on the linked strings by adding up the
contributions of the leptons and quarks and also taking into
account that the quarks come in 3 colours, that we have 2 loops
and $N_F$ families. The arithmetic gives the electric charge:
\begin{equation}
Q_A = 0 \ ;
\label{qa}
\end{equation}
the electric current along one of the strings:
\begin{equation}
J_A = 2e N_F \biggl ( 1 - {8 \over 3} sin^2\theta_W \biggr ) \ ;
\label{ja}
\end{equation}
the baryonic charge - in agreement with the results obtained by
indirect methods \cite{tvgf}:
\begin{equation}
B = 2 N_F cos(2\theta_W ) \ ;
\label{baryon}
\end{equation}
and the baryonic current along one of the strings:
\begin{equation}
J_B = - {{2N_F} \over 3} sin^2\theta_W \ .
\label{jb}
\end{equation}

\begin{table*}[hbt]
\setlength{\tabcolsep}{1.5pc}
\newlength{\digitwidth} \settowidth{\digitwidth}{\rm 0}
\catcode`?=\active \def?{\kern\digitwidth}
\caption{Summary of $Z-$, electric and baryon charges,
Fermi levels, energies, induced electric charges,
baryon numbers and electric and
baryonic currents
for the leptons and quarks. The charges $q_Z$ are for the
left-handed fermions. In evaluating the Fermi levels,
we have taken unit winding strings and
$\gamma \equiv sin^2 \theta_W \sim 0.23$.}
\label{tab:effluents}
\begin{tabular*}{\textwidth}{@{}l@{\extracolsep{\fill}}cccc}
\hline
                 & \multicolumn{1}{c}{$\nu_L$}
                 & \multicolumn{1}{c}{$e$}
                 & \multicolumn{1}{c}{$d$}
                 & \multicolumn{1}{c}{$u$}         \\
\hline
$2q_Z/\alpha$       &-1  &$    1-2\gamma$ &$1-{{2\gamma}\over 3}$
                                          &$-1+{{4\gamma}\over 3}$    \\
$q_A/e$             &0   &-1          & $-1 /3$
                                          & $2 / 3$     \\
$q_B$               &0   &0           &$1/ 3$
                                          & $1/ 3$      \\
$k_F^{(i)}$         &-1  &+1          & +1
                                          & -1              \\
$a E_i + {1 \over {24}}$
                    &$1\over 8$
                         &$    {1 \over 2}
                          \biggl ( 2\gamma -{1\over 2}\biggr )^2$
                                      &$    {1 \over 2}
          \biggl ( {{2\gamma}\over 3} - {1\over 2}\biggr )^2$
                                          &$    {1 \over 2}
          \biggl ( {{4\gamma}\over 3}- {1\over 2}\biggr )^2 $ \\
$Q_i$               &0   &$    \biggl ( 2\gamma -{1\over 2}\biggr )$
                                      &$    {1 \over 3}
          \biggl ( {{2\gamma}\over 3}- {1\over 2}\biggr )$
                                      &$    -{2 \over 3}
          \biggl ( {{4\gamma}\over 3}- {1\over 2}\biggr )$    \\
$B_i$               &0   &0           &$    -{1 \over 3}
          \biggl ( {{2\gamma}\over 3}- {1\over 2}\biggr )$
                                          &$    -{1 \over 3}
          \biggl ( {{4\gamma}\over 3}- {1\over 2}\biggr )$   \\
$j_i^A$             &0   &$-\biggl ( 2\gamma -{1\over 2}\biggr )$
                                      &$-{1 \over 3}
          \biggl ( {{2\gamma}\over 3}- {1\over 2}\biggr )$
                                      &$-{2 \over 3}
          \biggl ( {{4\gamma}\over 3}- {1\over 2}\biggr )$    \\
$j_i^B$             &0   &0           &${1 \over 3}
          \biggl ( {{2\gamma}\over 3}- {1\over 2}\biggr )$
                                          &$-{1 \over 3}
          \biggl ( {{4\gamma}\over 3}- {1\over 2}\biggr )$   \\
\hline
\multicolumn{5}{@{}p{120mm}}{}
\end{tabular*}
\end{table*}

It is interesting to work out the changes that occur when the
strings have linkage greater than 1. We have considered the situation when
a loop is threaded by $n$ other loops. In this case, the charges
$q_Z$ in Table 1 have to be multiplied by $n$ and then this changes the
Fermi levels. To analyze this situation, we begin
by writing all the quantities of interest in terms of the $K_i$
for a {\it single} loop with $N_F$ set equal to 1 for
convenience:
\begin{equation}
2 a E = K_\nu ^2 +K_e^2 + 3K_d^2 + 3 K_u^2
\label{gene}
\end{equation}
\begin{equation}
2 \Omega = K_\nu ^2 -K_e^2 - 3K_d^2 + 3 K_u^2
\label{twoom}
\end{equation}
\begin{equation}
Q_A/e = K_e + K_d + 2K_u
\label{qak}
\end{equation}
\begin{equation}
B = - K_d + K_u
\label{bak}
\end{equation}
\begin{equation}
J_A/e = - K_e - K_d + 2K_u
\label{jaak}
\end{equation}
\begin{equation}
J_B = K_d + K_u
\label{jbak}
\end{equation}
where the $K_i$ are defined in general in eqn. (\ref{ei}). For the
ground state, the Fermi levels are chosen to minimize the energy and
then the $K_i$ can be written as:
\begin{equation}
K_\nu (x) = \biggl \{ \biggl \{ {1 \over 2} \biggr \} \biggr \} \ ,
\label{knux}
\end{equation}
\begin{equation}
K_e (x) = \biggl \{ \biggl \{ 3x - {1 \over 2} \biggr \} \biggr \} \ ,
\label{kex}
\end{equation}
\begin{equation}
K_d (x) = \biggl \{ \biggl \{ x - {1 \over 2} \biggr \} \biggr \} \ ,
\label{kdx}
\end{equation}
and,
\begin{equation}
K_u (x) = - \biggl \{ \biggl \{ 2x - {1 \over 2} \biggr \} \biggr \} \
\label{kux}
\end{equation}
where $x=2n\gamma/3$ and
$\{ \{ y \} \}$ is the smallest (in magnitude) fractional part
of  $y$.
(For example, for $y=+0.2\ {\rm and} \ +0.7$,
these are 0.2 and -0.3. For half-integral values of
$y$, there is a sign ambiguity but this does not affect our results
for the energy and angular momentum which are quadratic in $K_i$.
The charges and currents can depend on the choice of sign but only
at isolated
points. For irrational values of $sin^2\theta_W$, $x$ can never
take on these values.
To find $\{ \{ y \} \}$ for negative $y$ we can use the property
$ \{ \{ - y \} \} = - \{ \{ y \} \}$.)

It is easy to check that
\begin{equation}
K_i (x+k) = K_i (x) \ , \ \ \ k \in {\cal Z}
\label{kiper}
\end{equation}
and so we can restrict our attention to $x \in (0,1)$. In this
interval, we can check that,
\begin{equation}
\biggl \{ \biggl \{ mx - {1 \over 2} \biggr \} \biggr \} =
mx - {1 \over 2} - i \ , \ \ \ x \in ( {i\over m}, {{i+1} \over m} )
\label{genfor}
\end{equation}
where, $m$ is an integer and $i = 0,1,...,m-1$. This general formula
yields:
\begin{equation}
K_i (1-x) = - K_i(x)
\label{kionex}
\end{equation}
and so we need only calculate the quantities of interest in the interval
$(0, 1/2)$. Outside this range, we can write down the values of the
quantities by using the symmetry in (\ref{kionex}).


In Table 2 we show the expressions for the energy, angular momentum,
charges and currents for one loop threaded by $n$ strings in terms of
$x = 2n\gamma /3$. Note that the angular momentum is non-vanishing
only when the electric charge is non-zero, and
as defined by $\Omega$, does not come in half-integral units. This
is of no concern because $\Omega$ does not include the angular
momentum of the bosonic background and so can have any fractional
value.  On the other hand,
the generalized angular momentum found in eqn. (\ref{totgam}) is
always half-integral as can be seen, for example, by expanding
out the right-hand side of (\ref{totgam}).
As shown in Table 2, in the ground
state, the baryon number of the single loop is given by
$nN_F cos2\theta_W$ up to an additive integer. This is also
what we expect from integrating out the anomaly equation\cite{tvgf}.

\begin{table*}[hbt]
\setlength{\tabcolsep}{.99pc}
\catcode`?=\active \def?{\kern\digitwidth}
\caption{Expressions for the energy, angular momentum, charges
and currents in terms of $x =  2n\gamma /3$. We have omitted the
multiplicative factor $N_F$ in all the expressions for convenience.
}
\label{tab:results}
\begin{tabular*}{\textwidth}{@{}l@{\extracolsep{\fill}}cccc}
\hline
                 & \multicolumn{1}{c}{$x \in (0,1/3)$}
                 & \multicolumn{1}{c}{$(1/3,1/2)$}
                 & \multicolumn{1}{c}{$(1/2,2/3)$}
                 & \multicolumn{1}{c}{$(2/3,1)$}         \\
\hline
$aE$       &$12x^2-6x+1$  &$12x^2-9x+2$  &$12x^2-15x+5$
                                          &$12x^2-18x+7$    \\
$\Omega$             &0   &$3x-1$& $-3x+2$ & 0     \\
$Q_A /e$               &0   &-1           & +1 & 0      \\
$B$         &$-3x+1$  &$-3x+1$   &$-3x+2$ &$-3x+2$   \\
$J_A /e$    &$-8x+2$  &$-8x+3$   &$-8x+5$  &$-8x+6$ \\
$J_B$       &$-x$     &$-x$      &$1-x$   &$1-x$   \\
\hline
\multicolumn{5}{@{}p{120mm}}{}
\end{tabular*}
\end{table*}

The energy of the fermionic ground state shows a complicated
dependence on $x$ as is demonstrated in Fig. 2. By changing
the linking number, we can change $x$ in discrete steps and
sample different points on the $E(x)$ curve. Since $E(x)$ does
not have a monotonic dependence on $x$, the energy of strings
that are linked $n$ times bears no simple relation to those
linked $m$ times. In particular, the energy does not continue
to decrease as we consider strings that have higher linkage.
The lowest energy possible is when $x=1/4$ but, if $\gamma$
is irrational, there will not be any value of the linkage for
which $x$ will be $1/4$. (Though we can choose values of $n$
for which $x$ is arbitrarily close to $1/4$).

The curve in Fig. 2 has the interpretation of a potential for
the quantity $x$ and tells us which values of $x$ are most
favoured. In a physical situation, however, we should calculate
the free energy which would be large for large linkage.
This would remove the degeneracy between minima of different
linkage and, for high enough string density, we would probably be
left with two degenerate global minima - one
occurring at some positive value of $x$ and another
at the negative of this value.

\

\

\

\section{Conclusions}

In this paper we have studied boson and fermion zero modes
on linked string configurations including the case when the
boson or fermion has an Aharanov-Bohm interaction with the
string.
{}From this analysis we can calculate the various
charges on the linked strings. We confirm that the baryon
number on linked loops of electroweak  string is that
found in Ref. \cite{tvgf}: $2N_F cos(2\theta_W )$. In addition,
our direct calculations enable us to calculate the electric
charge, electric current
and electomagnetic current on the linked strings. The charge
vanishes on singly linked unit winding string loops
but the electric and baryonic currents are non-vanishing.
For strings with higher linking numbers, the situation gets
more complicated. While the energy of linked strings is always
less than the energy of unlinked strings, the energy is not
a monotonic function of the linking number and fluctuates
as shown in Fig. 2. In the bosonic case, we also
find that the ground state of the linked configuration is
current carrying.

We have found a ``super-Bogomolnyi'' limit
in which the Dirac equations for the fermions can be reduced
by one order to an algebraic equation. This is the limit
when the scalar, vector and fermion masses are all equal
and the charge of the left-handed fermion vanishes.
Probably this limit and the result are related to a manifestation
of supersymmetry but we are not aware of a discussion of
this point in the literature.

Finally, we propose that there may be a gravitational
analogue of the system of linked strings. This is because
a particle moving in a conical space - such as produced
by a cosmic string - has a
gravitational Aharanov-Bohm interaction with the line
source that produces the conical space. Then if
we consider two circular line sources each of fixed
radius $a$, the total source length in the absence of
linkage is simply $4\pi a$. But if the sources are linked,
the gravitational Aharanov-Bohm effect reduces the
total length of the sources to $2(2\pi - \delta ) a$
where $\delta$ is the conical deficit angle produced
by the sources. So the linked sources have less length
for the same radius as compared to the unlinked
sources. This is the gravitational analogue of the result
that the ground state of linked strings is lower than
the ground state of unlinked strings.

Even though there are traveling wave solutions (``zero
modes'') on gravitating strings \cite{tvdg}, there does not seem
to be any gravitational analogue of the currents
that run along linked strings - at least, in the context
of static, conical spaces in conventional general relativity.
This is because there is nothing to distinguish one direction
along the line source over the other. However, it may be
that there are non-vanishing currents along the sources if the
linked sources are spinning \cite{pl}. We feel that this system
is worth exploring further.

\

\

\

\section*{Acknowledgements}

We are grateful to  the steady stream of visitors at the Isaac
Newton Institute for advice and suggestions. In particular,
we would like to thank
Tom Kibble, Patrick  Peter, Rich Holman, Mark Hindmarsh,
Rob Brandenberger, Roman Jackiw, Patricio Letelier, Hector
de Vega and especially Nick Manton for crucial advice at critical
times. JG is grateful to the SERC for support under
Grant No. 15091-AOZ-L9 at DAMTP, University of Cambridge where most
of this work was done. TV is grateful to the Rosenbaum Foundation
for supporting his stay at the Isaac Newton Institute.

\section*{Figure captions}

\begin{itemize}

\item{\bf Fig. 1} A pair of linked loops.

\item{\bf Fig. 2} The fermionic ground-state energy $E$ (in units
of $1/a$) of
a single loop of string threaded by $n$ other strings versus
$x = (2n/3)sin^2\theta_W$.

\end{itemize}


\begin{thebibliography}{999}

\bibitem{rjcr} R. Jackiw and C. Rebbi, Phys. Rev. {\bf D13}, 3398 (1976).

\bibitem{jgfw} J. Goldstone and F. Wilczek, Phys. Rev. Lett.
{\bf 47}, 986 (1981).

\bibitem{wi85} E. Witten, Nucl. Phys. {\bf B249}, 557 (1985).

\bibitem{yn} Y. Nambu, Nucl. Phys. {\bf B130}, 505 (1977).

\bibitem{nm} N. S. Manton, Phys. Rev. {\bf D28}, 2019 (1983).

\bibitem{tv1}  T. Vachaspati, Phys. Rev. Lett. {\bf 68}, 1977 (1992);
{\bf 69}, 216(E) (1992); Nucl. Phys. {\bf B397}, 648 (1993).

\bibitem{mewp} M. Earnshaw and W. Perkins, Phys. Lett. {\bf B328},
337 (1994).

\bibitem{tvgf} T. Vachaspati and G. B. Field, Phys. Rev. Lett. {\bf 73},
373 (1994).

\bibitem{manton} N. S. Manton, Ann. of Phys. {\bf 159}, 220 (1985).

\bibitem{bogo} E. B. Bogomolnyi, Sov. J. Nucl. Phys. {\bf 24}, 449 (1976).

\bibitem{chengli} T. P. Cheng and L. F. Li, ``Gauge Theory of
Elementary Particle Physics'', Oxford University Press (1991).

\bibitem{tvmb} T. Vachaspati and M. Barriola, Phys. Rev. Lett. {\bf 69},
1867 (1992).

\bibitem{mbtvmb} M. Barriola, T. Vachaspati and M. Bucher, Phys. Rev.
{\bf D50}, 2819 (1994).

\bibitem{ew} E. Weinberg, Phys. Rev. {\bf D24}, 2669 (1981).

\bibitem{pvsf} P. di Vecchia and S. Ferrara, Nucl. Phys. {\bf B130},
93 (1977).

\bibitem{hnpo} H. B. Nielsen and P. Olesen, Nucl. Phys. {\bf B61},
45 (1973).

\bibitem{gape94} J. Garriga and
P. Peter, Class. Quant. Grav. {\bf 11}, 1743 (1994).

\bibitem{grry} I.S. Gradshteyn and I.M. Ryzhik, Tables of integrals,
series and products, Academic Press, New York (1980).

\bibitem{hdv} H. de Vega, Phys. Rev. D{\bf 18}, 2932 (1978).

\bibitem{mjlptv} M. James, L. Perivolaropoulos and T. Vachaspati,
Nucl. Phys. {\bf B395}, 534 (1993).

\bibitem{tvdg} D. Garfinkle and T. Vachaspati, Phys. Rev. D{\bf 42},
1960 (1990).

\bibitem{pl} P.S. Letelier, Isaac Newton Institute preprint; to
appear in Classical and Quantum Gravity (1994).

\end{thebibliography}
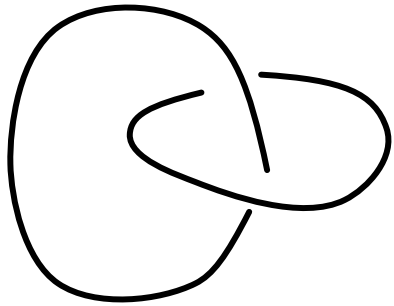
\end{document}